\documentclass[twocolumn,10pt]{article}

\usepackage[a4paper,margin=16mm]{geometry}
\usepackage{graphicx}
\usepackage{booktabs}
\usepackage{multirow}
\usepackage{authblk}
\usepackage{amsmath}
\usepackage{siunitx}
\usepackage[dvipsnames]{xcolor}
\usepackage{caption}
\PassOptionsToPackage{hyphens}{url}
\usepackage[hidelinks]{hyperref}
\usepackage[small]{titlesec}
\usepackage{tikz}
\usepackage{pgfplots}
\usepackage{pgfplotstable}
\pgfplotsset{compat=1.18}
\usepgfplotslibrary{fillbetween}
\usetikzlibrary{patterns}
\usetikzlibrary{positioning,arrows.meta,fit,calc,shapes.geometric}
\usepackage[round,authoryear,sort]{natbib}
\usepackage[capitalise,nameinlink]{cleveref}

\definecolor{cbGray}{RGB}{153,153,153}
\definecolor{cbVermillion}{RGB}{213,94,0}
\definecolor{cbTeal}{RGB}{0,158,115}

\titlespacing*{\section}{0pt}{2ex}{1ex}
\titlespacing*{\subsection}{0pt}{1.8ex}{0.9ex}
\titlespacing*{\subsubsection}{0pt}{1.6ex}{0.8ex}
\titlespacing*{\paragraph}{0pt}{1.4ex}{1em} 

\title{\vspace{10pt}\bfseries AIFS-TC: A simple correction competitive with the operational frontier for tropical cyclone intensity forecasting }

\author[1$\dagger$*]{Anna Allen}
\author[1$\dagger$]{Wessel P.\ Bruinsma}
\author[2]{Michael Maier-Gerber}
\author[2]{Harrison Cook}
\author[2]{Matthew Chantry}
\author[1]{Richard E.\ Turner}
\affil[1]{University of Cambridge}
\affil[2]{European Centre for Medium-Range Weather Forecasts}
\affil[*]{Correspondence: A.\ Allen anna@brightband.com}
\affil[$\dagger$]{These authors contributed equally.}
\date{}

\begin{document}
\maketitle

\begin{abstract}
\noindent 
AI weather models are in the process of revolutionising weather forecasting. While these models have been shown to achieve superior performance to physics-based NWP in forecasting tropical cyclone (TC) tracks, they tend to dramatically underestimate intensity. Here we present AIFS-TC, a simple correction to the AIFS-Single model that is competitive with the operational state-of-the-art for forecasting maximum wind speed and minimum central pressure at lead times of \SI{12}{h} to seven days. This performance also holds for rapid intensification events. Notably, the entire system was autonomously designed and built by a large language model (Claude~Fable~5) in a few hours, directed through a small number of natural-language prompts by a single domain scientist. That the operational frontier can be reached with an open-source AI forecast model (AIFS-Single) and relatively simple, cheap post-processing is significant for TC science, and points to agentic coding as a route to rapid exploration and progress in life-saving early-warning systems in other domains.

\end{abstract}

\section{Introduction}
The last three years have redrawn the map of weather prediction. Machine-learning models (GraphCast,
GenCast, Aurora, ECMWF's own AIFS) now match or beat operational physics-based model for medium-range fields at a
fraction of the running cost \citep{graphcast,gencast,aurora,aifs}. Despite these advances, forecasting TCs has remained challenging. AI models outperform operational predictions for TC track \citep{aurora}, but tend to systematically underestimate intensity \citep{demaria_aiwp}.

In recognition of this problem, various attempts have been made to improve forecasts of TC intensity using AI methods. Many of them post-process existing global machine-learning models such as AIFS  \citep{gomez2025global,tcbench,baguancyclone,istm,fuxitc,hits}. 
These systems demonstrate that post-processing can provide
substantial improvements over raw model output,
but stop short of a comparison against state-of-the-art systems.
To date the only AI model with similar performance to the state-of-the-art is Google's FNV3 \citep{FNV3}. Unlike the cheap post-processing methods above, this is a bespoke forecasting model specifically trained for TCs.

Reaching this frontier has, until now, required a specialist team, bespoke
\looseness=-1
pipelines and a substantial compute budget. We ask a different question: how much resource is really necessary to attain state-of-the-art performance? We show that a freely available open weather model, ECMWF's AIFS-Single \citep[referred to hereon as AIFS;][]{aifs}, combined with a deliberately simple and low-cost learned post-processing step, reaches the level of the operational state-of-the-art currently set by both FNV3 and the National Hurricane Center (NHC) official forecast.

\begin{figure*}[t]
\centering
\pgfplotstableread[col sep=comma]{
name,numstorms,aifsmean,aifslower,aifsupper,predmean,predlower,predupper,fnv3mean,fnv3lower,fnv3upper,ofclmean,ofcllower,ofclupper
Global,70,28.85,23.49,34.32,10.75,9.32,12.18,10.18,8.59,11.75,nan,nan,nan
NAEP,29,35.31,27.34,41.60,11.05,9.34,12.44,10.72,8.52,12.92,10.70,9.04,12.25
RI,11,71.06,59.75,78.17,20.92,14.82,29.33,23.78,18.48,31.07,20.11,15.82,25.10
NA,10,39.69,28.18,45.70,11.66,8.58,13.61,10.65,7.95,12.33,11.88,9.17,14.51
EP,19,31.78,23.19,41.98,10.55,8.63,12.53,10.77,7.78,14.46,9.76,7.97,11.60
WP,28,22.42,16.60,28.91,9.79,7.67,12.08,9.04,6.99,11.08,nan,nan,nan
NI,5,10.94,5.39,24.96,4.74,2.35,7.42,4.62,3.42,7.78,nan,nan,nan
SH,8,24.96,9.77,26.34,10.70,3.51,11.52,11.37,4.54,12.13,nan,nan,nan
}\results
 
\begin{tikzpicture}
\begin{axis}[
    set layers, 
    width=\linewidth, height=5.5cm,
    title style={font=\small\bfseries, yshift=1pt},
    ybar=1.5pt,
    bar width=10pt,
    ymin=0,
    ymax=75,
    ytick distance=10,
    enlarge y limits={upper, value=0.08},
    enlarge x limits=0.10,
    ylabel={Maximum wind speed MAE (kt)},
    symbolic x coords={
        Global,
        NAEP,
        RI,
        NA,
        EP,
        WP,
        NI,
        SH,
    },
    xticklabels={
        {\bfseries Global \\ \scriptsize (70 storms)},
        {\bfseries NA \& EP \\ \scriptsize (29 storms)},
        {\bfseries Rap.\ int. \\ \scriptsize (11 storms)},
        {NA \\ \scriptsize (10 storms)},
        {EP \\ \scriptsize (19 storms)},
        {WP \\ \scriptsize (28 storms)},
        {NI \\ \scriptsize (5 storms)},
        {SH \\ \scriptsize (8 storms)},
    },
    x tick label style={align=center},
    xtick=data,
    ymajorgrids,
    grid style={line width=.4pt, draw=gray!25},
    axis line style={line width=.5pt, draw=gray!60},
    tick style={draw=none},
    tick label style={font=\small},
    label style={font=\small},
    legend style={
    at={(1,1)}, anchor=north east,
    xshift=-0.25cm, yshift=-0.25cm,
    legend columns=-1, draw=gray!60, fill=white, font=\small,
    /tikz/every even column/.append style={column sep=0.9em},
    },
    legend image code/.code={\draw[#1] (0cm,-0.1cm) rectangle (0.28cm,0.14cm);},
    error bars/y dir=both,
    error bars/y explicit,
    error bars/error bar style={line width=.5pt, black},
    error bars/error mark options={rotate=90, mark size=1.7pt, line width=.5pt, black},
]
\addplot[fill=white, draw=black] table[
    x=name, y=aifsmean,
    y error plus  expr=\thisrow{aifsupper}-\thisrow{aifsmean},
    y error minus expr=\thisrow{aifsmean}-\thisrow{aifslower},
  ] {\results};
\addplot[fill=black!80, draw=black] table[
    x=name, y=predmean,
    y error plus  expr=\thisrow{predupper}-\thisrow{predmean},
    y error minus expr=\thisrow{predmean}-\thisrow{predlower},
  ] {\results};
\addplot[fill=gray!50, draw=black] table[
    x=name, y=fnv3mean,
    y error plus  expr=\thisrow{fnv3upper}-\thisrow{fnv3mean},
    y error minus expr=\thisrow{fnv3mean}-\thisrow{fnv3lower},
  ] {\results};
\addplot[pattern=north east lines, draw=black] table[
    x=name, y=ofclmean,
    y error plus  expr=\thisrow{ofclupper}-\thisrow{ofclmean},
    y error minus expr=\thisrow{ofclmean}-\thisrow{ofcllower},
  ] {\results};
\coordinate (lowerleft) at (rel axis cs:0,0);
\coordinate (upperleft) at (rel axis cs:0,1);
\path ($(lowerleft)!.3!(axis cs:Global,0)$) coordinate (leftmid);
\path ($(axis cs:RI,0)!.5!(axis cs:NA,0)$) coordinate (rightmid);
\begin{pgfonlayer}{axis background}
  \fill[gray!9, draw=gray!60] (leftmid) rectangle (rightmid |- upperleft);
\end{pgfonlayer}
\legend{AIFS, AIFS-TC, FNV3, OFCL}
\end{axis}
\end{tikzpicture}
\caption{
    \textbf{AIFS-TC is competitive with the operational frontier.}
    Shows maximum wind speed MAE averaged over lead times 12--168 hours on the held-out year 2025 for all storms globally, for cases of rapid intensification (``Rap.\ int.''), and for various basins.
    NA stands for the North Atlantic; EP for the East Pacific; WP for the West Pacific; NI for the North Indian; and SH for the Southern Hemisphere.
    For each group, numbers are computed from the subset of data points for which forecasts are available for all models.
    In particular, the numbers for rapid intensification are restricted to the North Atlantic and East Pacific to enable a comparison with OFCL.
    Error bars indicate 95\% confidence intervals, which are computed by resampling entire storms with replacement and taking the 2.5\% and 97.5\% quantiles.
}
\label{fig:headline}
\end{figure*}
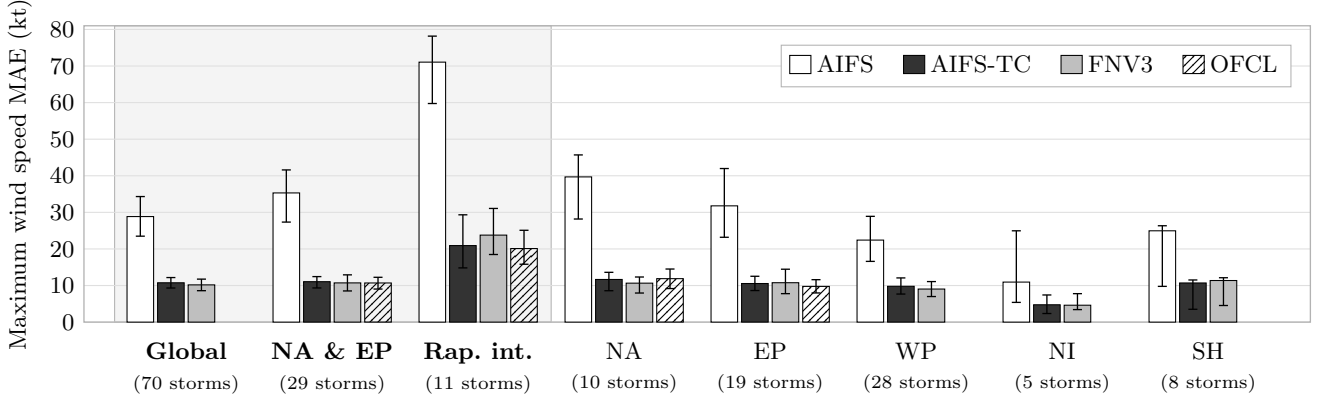
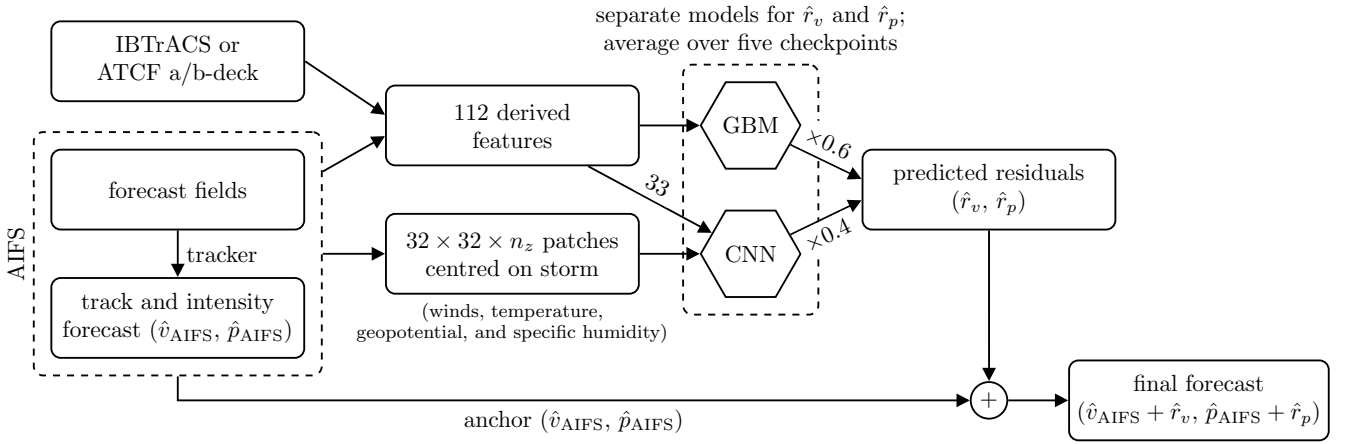
\begin{figure*}[t]
    \centering
    \scalebox{0.88}{\begin{tikzpicture}[
        >={Triangle[length=2mm, width=2mm]},
        thick,
        node distance = 7mm and 9mm,
        box/.style = {
            draw,
            rounded corners,
            align=center,
            minimum width=38mm,
            minimum height=12mm,
        },
        model/.style = {
            draw,
            regular polygon,
            regular polygon sides=6,
            xscale=1,
            align=center,
            minimum width=15mm,
            minimum height=12mm,
            inner sep=1pt,
        },
        sum/.style = {
            draw,
            circle,
            inner sep=1pt,
            minimum size=5.5mm,
        },
        scale=0.8,
    ]
    
    \node[box] (ab) {IBTrACS or \\ATCF a/b-deck};
     
    \node[box, below=of ab] (ff) {forecast fields};
    \node[box, below=of ff] (tf)
        {track and intensity \\forecast $(\hat v_{\mathrm{AIFS}}$, $\hat p_{\mathrm{AIFS}})$};
    \node[draw, dashed, rounded corners, inner sep=2.5mm, fit=(ff)(tf)] (aifs) {};
    \node[anchor=south,rotate=90] at (aifs.west) {AIFS};

    \path ($(ab.east)!0.5!(ff.east)$) coordinate (mid1);
    \path ($(ff.east)!0.5!(tf.east)$) coordinate (mid2);
     
    \node[box, right=12mm of mid1] (feat)  {112 derived\\ features};
    \node[box, right=12mm of mid2] (patch) {$32\times32\times n_z$ patches \\ centred on storm};
    \node[anchor=north, align=center] () at (patch.south) {\footnotesize (winds, temperature, \\[-3pt] \footnotesize geopotential, and specific humidity)};
         
    \node[model, right=of feat]  (gbm) {};
    \node () at (gbm) {GBM};
    \node[model, right=of patch] (cnn) {};
    \node () at (cnn) {CNN};
    \node[box, right=of $(gbm.east)!0.5!(cnn.east)$] (blend) {predicted residuals \\ $(\hat r_v$, $\hat r_p)$};
    \node[draw, dashed, rounded corners, inner sep=2.5mm, fit=(gbm)(cnn)] (models) {};
    \node[anchor=south, align=center] at (models.north) {separate models for $\hat r_v$ and $\hat r_p$; \\ average over five checkpoints};
     
    \node[sum] (plus) at ($(blend |- aifs.south) + (0,-4.5mm)$) {$+$};
    \node[box, right=of plus] (out) {final forecast\\ $(\hat v_{\mathrm{AIFS}}+\hat r_v$, $\hat p_{\mathrm{AIFS}} +\hat r_p)$};
     
    \draw[->] (ff) -- (tf) node [pos=0.5, anchor=west] {tracker};
    \draw[->] (ab.east) -- ([yshift=2mm]feat.west);
    \draw[->] (aifs)  -- ([yshift=-2mm]feat.west);
    \draw[->] (aifs.east |- patch) -- (patch);
    \draw[->] (feat)  -- (gbm);
    \draw[->] (feat)  -- node[above, sloped, pos=0.5, fill=white] {33} (cnn);
    \draw[->] (patch) -- (cnn);
    \draw[->] (gbm) -- node[above, sloped, pos=0.42, fill=white] {$\times 0.6$} ([yshift=2mm]blend.west);
    \draw[->] (cnn) -- node[below, sloped, pos=0.42, fill=white] {$\times 0.4$} ([yshift=-2mm]blend.west);
    \draw[->] (blend) -- node[right, pos=0.4] {} (plus);
    \draw[->] (aifs.south) |- node[below, pos=0.75]
              {anchor $(\hat v_{\mathrm{AIFS}}$, $\hat p_{\mathrm{AIFS}})$} (plus);
    \draw[->] (plus) -- (out);
     
    \end{tikzpicture}}

\caption{
    \textbf{The architecture of AIFS-TC.}
    AIFS-TC learns an additive correction to the AIFS-Single intensity forecast (the ``anchor'').
    The additive correction is produced
    by a $0.6$/$0.4$ blend of (1) a gradient-boosted tree ensemble on 112 derived features; and (2) a convolutional neural network on 33 of these 112 features and 3D patches of AIFS forecast fields centred on the storm.
    The patches have $n_z=4$ by selecting pressure levels \SI{850}{hPa}, \SI{700}{hPa}, \SI{500}{hPa}, and \SI{200}{hPa}.
    The final maximum wind speed forecast is clipped to $[\text{\SI{5}{kt}},\text{\SI{200}{kt}}]$ and the final minimum central pressure forecast to $[\text{\SI{850}{mb}},\text{\SI{1025}{mb}}]$.
}
\label{fig:schematic}
\end{figure*}

\section{AIFS-TC}
\label{sec:aifs-tc}
Our approach, named AIFS-TC, learns an additive correction to AIFS intensity forecasts by modelling AIFS intensity forecast errors.
The additive correction is produced from the full AIFS forecast, which includes
forecasted three-dimensional fields (winds, temperature, geopotential, and specific humidity) cropped to patches centred on the storm
and predictions for track and intensity (which are produced from a tracker, see \citet{tctrackerecmwf}, applied to the AIFS).
\looseness=-1
The model is trained on TCs and AIFS hindcasts for years 2016--2024.
The year 2025 is held out during development and used as the test set.

To build this system, the first author specified in natural language the scientific goal, the data that could be used, and the way skill would be judged; the coding agent did everything else. AIFS forecasts were made available to the agent, which independently downloaded ground truth from the International Best Track Archive for Climate Stewardship \citep[IBTrACS;][]{ibtracks1,ibtracks2}; designed and engineered 112 features; implemented and trained a model; and ran the evaluation. The human contribution was problem specification and rigorous validation of the outputs. There was a general tendency for Claude Fable 5 to find solutions that were overcomplicated. The agent was then directed to simplify its first solution through further interactions. We found that the simplification did not compromise performance, leaving the model and results reported here.

The resulting approach produces predictions for maximum sustained \SI{10}{m} wind speed and the minimum central pressure by applying an additive correction to AIFS's intensity forecasts. This additive correction is obtained by averaging the output of five gradient boosting machines (GBMs), averaging the output of five convolutional neural networks (CNNs), and combining these two averages with weights $0.6$ and $0.4$ respectively (\cref{fig:schematic}).
The models are separately trained for maximum wind speed and minimum central pressure, meaning that there are $|\{v, p\} \times \{\text{GBM}, \text{CNN}\} \times \{1, \ldots, 5\}| = 20$ checkpoints in total.
The five checkpoints for each GBM and CNN come from a five-fold cross-validation split:
storms in the training data were partitioned into five approximately equal groups, and each group was held out once with the model trained on the other four.
\looseness=-1
During training, the loss for rapid-intensification cases was up-weighted by a factor two to reflect their practical importance.
The inputs to the GBMs and CNNs are derived from the AIFS forecast itself and the operational best track from the ATCF b-deck, allowing AIFS-TC to run the moment a forecast is delivered. 

We evaluate against the IBTrACS on the held-out test year 2025. We measure the mean absolute error (MAE) and mean bias for maximum wind speed in knots averaged over lead times 12--168~hours. We consider performance against unprocessed AIFS forecasts, the ensemble mean of the operational version of FNV3, and official forecasts (OFCL) from the NHC (which only exist for the North Atlantic and East Pacific). We note that FNV3 is an ensemble model and that official forecasts are produced by human forecasters guided by an ensemble of dynamical models. In contrast, for simplicity, AIFS-TC is derived from AIFS-Single, a single, deterministic product, along with a small amount of operationally available TC information.
We also note that OFCL should be compared against early models, for which forecasts are available within three hours after synoptic (initialisation) time.
AIFS-TC and FNV3 are both late models, so the evaluation slightly disadvantages OFCL.
We have retrained AIFS-TC with a \SI{12}{h} delay and our results do not exhibit significant changes, including on rapid intensification.
We leave an early version of AIFS-TC with a \SI{6}{h} delay for future work.

\begin{table*}[t]
\centering
\caption{\textbf{Intensity forecast error.}
Shows the MAE and mean bias for maximum wind speed and minimum central pressure averaged over lead times 12--168 hours on the held-out year 2025.
For each row, numbers are computed from the subset of data points for which forecasts are available for all models.
In particular, the numbers for rapid intensification maximum wind speed are restricted to the North Atlantic and East Pacific to enable a comparison with OFCL.
The best number (lowest MAE or bias closest to zero) and any other number not statistically significantly different from the best number are boldfaced.
Entire storms are resampled with replacement, and a difference is deemed significant if zero lies outside the 2.5\%--97.5\% quantile interval of the difference in the metric (for the bias, the difference in magnitude).
}
\label{tab:main}
\small
\addtolength{\tabcolsep}{-3.5pt}
\hfill
\begin{tabular}{
    @{}
    l
    >{\raggedleft\arraybackslash}p{1cm}
    >{\raggedleft\arraybackslash}p{1cm}
    >{\raggedleft\arraybackslash}p{1cm}
    >{\raggedleft\arraybackslash}p{1cm}
    @{}
}
\toprule
\multirow{2}{*}{MAE}
    & \multirow{2}{*}{AIFS}
    & AIFS\newline-TC
    & \multirow{2}{*}{FNV3}
    & \multirow{2}{*}{OFCL} \\
\midrule
\multicolumn{5}{@{}l}{\textsc{Maximum wind speed (kt)}}\\
\quad Global & 28.9 & \textbf{10.8} & \textbf{10.2} & \\
\quad N.\ Atl. \& E.\ Pac. & 35.3 & \textbf{11.0} & \textbf{10.7} & \textbf{10.7} \\
\quad Rapid intensification & 71.1 & \textbf{20.9} & \textbf{23.8} & \textbf{20.1} \\
\addlinespace[3pt]\multicolumn{5}{@{}l}{\quad\textsc{By basin:}}\\
\quad\quad North Atlantic & 39.7 & \textbf{11.7} & \textbf{10.6} & \textbf{11.9} \\
\quad\quad East Pacific & 31.8 & \textbf{10.6} & \textbf{10.8} & \textbf{9.8} \\
\quad\quad West Pacific & 22.4 & \textbf{9.8} & \textbf{9.0} & \\
\quad\quad North Indian & 10.9 & \textbf{4.7} & \textbf{4.6} & \\
\quad\quad Southern Hemisphere & 25.0 & \textbf{10.7} & 11.4 & \\
\addlinespace[5pt]
\multicolumn{5}{@{}l}{\textsc{Minimum central pressure (mb)}}\\
\quad Global & 11.6 & \textbf{7.9} & \textbf{7.8} & \\
\quad N.\ Atl. \& E.\ Pac. & 15.4 & \textbf{8.8} & \textbf{8.6} & \\
\quad Rapid intensification & 38.0 & \textbf{18.8} & \textbf{20.8} & \\
\addlinespace[3pt]\multicolumn{5}{@{}l}{\quad\textsc{By basin:}}\\
\quad\quad North Atlantic & 19.2 & \textbf{10.7} & \textbf{10.8} & \\
\quad\quad East Pacific & 12.2 & \textbf{7.2} & \textbf{6.7} & \\
\quad\quad West Pacific & \textbf{8.4} & \textbf{7.0} & \textbf{7.0} & \\
\quad\quad North Indian & \textbf{3.7} & \textbf{3.1} & \textbf{2.9} & \\
\quad\quad Southern Hemisphere & \textbf{10.3} & \textbf{8.6} & 9.1 & \\
\bottomrule
\end{tabular}
\hspace{15pt}
\begin{tabular}{
    @{}
    l
    >{\raggedleft\arraybackslash}p{1cm}
    >{\raggedleft\arraybackslash}p{1cm}
    >{\raggedleft\arraybackslash}p{1cm}
    >{\raggedleft\arraybackslash}p{1cm}
    @{}
}
\toprule
\multirow{2}{*}{Bias}
    & \multirow{2}{*}{AIFS}
    & AIFS\newline-TC
    & \multirow{2}{*}{FNV3}
    & \multirow{2}{*}{OFCL} \\
\midrule
\multicolumn{5}{@{}l}{\textsc{Maximum wind speed (kt)}}\\
\quad Global & -28.8 & \textbf{-1.9} & -4.2 & \\
\quad N.\ Atl. \& E.\ Pac. & -35.3 & \textbf{-3.2} & \textbf{-4.5} & \textbf{0.6} \\
\quad Rapid intensification & -71.1 & \textbf{-19.5} & \textbf{-23.0} & \textbf{-19.7} \\
\addlinespace[3pt]\multicolumn{5}{@{}l}{\quad\textsc{By basin:}}\\
\quad\quad North Atlantic & -39.7 & \textbf{-4.0} & \textbf{-3.6} & \textbf{0.2} \\
\quad\quad East Pacific & -31.8 & \textbf{-2.5} & \textbf{-5.3} & \textbf{0.9} \\
\quad\quad West Pacific & -22.3 & \textbf{1.5} & \textbf{-2.0} & \\
\quad\quad North Indian & -10.9 & \textbf{0.1} & \textbf{-0.9} & \\
\quad\quad Southern Hemisphere & -24.9 & \textbf{-9.7} & -11.0 & \\
\addlinespace[5pt]
\multicolumn{5}{@{}l}{\textsc{Minimum central pressure (mb)}}\\
\quad Global & 10.2 & \textbf{1.5} & 3.4 & \\
\quad N.\ Atl. \& E.\ Pac. & 14.4 & \textbf{3.0} & 4.7 & \\
\quad Rapid intensification & 38.0 & \textbf{17.2} & \textbf{18.6} & \\
\addlinespace[3pt]\multicolumn{5}{@{}l}{\quad\textsc{By basin:}}\\
\quad\quad North Atlantic & 18.8 & \textbf{5.4} & \textbf{6.3} & \\
\quad\quad East Pacific & 10.7 & \textbf{1.0} & \textbf{3.2} & \\
\quad\quad West Pacific & 6.6 & \textbf{-1.1} & \textbf{1.3} & \\
\quad\quad North Indian & \textbf{2.4} & \textbf{-0.3} & \textbf{-0.1} & \\
\quad\quad Southern Hemisphere & 9.6 & \textbf{7.7} & \textbf{8.7} & \\
\bottomrule
\end{tabular}
\hfill\strut
\end{table*}

\begin{figure*}[t]
\centering
\pgfplotstableread[col sep=comma]{
cat,aifsmean,aifslower,aifsupper,predmean,predlower,predupper,fnv3mean,fnv3lower,fnv3upper,ofclmean,ofcllower,ofclupper,n,nstorm
TD,6.56,4.54,9.17,6.94,4.24,10.34,5.30,2.72,7.85,7.01,4.18,10.73,77.00,12.00
TS,19.81,17.65,21.82,7.75,6.57,8.96,6.31,5.10,7.83,7.74,6.53,9.51,605.00,29.00
Cat1,34.06,30.63,37.80,9.77,7.68,12.06,8.94,6.82,11.82,10.28,7.30,13.83,304.00,16.00
Cat2,43.73,36.91,55.02,10.92,8.36,15.10,11.68,8.41,17.26,10.88,7.47,13.23,153.00,7.00
Cat3,65.16,57.73,71.36,15.36,9.35,22.06,19.13,10.81,27.40,14.78,11.59,19.50,113.00,8.00
Cat4,79.02,72.63,84.65,25.69,16.44,34.41,26.63,16.77,36.95,20.85,15.70,26.90,117.00,6.00
Cat5,104.06,88.16,109.31,27.28,24.56,30.61,31.56,27.61,38.22,27.34,19.17,38.33,32.00,2.00
}\results
 
\begin{tikzpicture}
\begin{axis}[
    set layers, 
    width=\linewidth, height=5cm,
    title style={font=\small\bfseries, yshift=1pt},
    ybar=1.5pt,
    bar width=10pt,
    ymin=0,
    ymax=105,
    ytick distance=20,
    minor y tick num=1,
    enlarge y limits={upper, value=0.08},
    enlarge x limits=0.10,
    ylabel={Intensity MAE (kt)},
    symbolic x coords={
        TD,
        TS,
        Cat1,
        Cat2,
        Cat3,
        Cat4,
        Cat5,
    },
    xticklabels={
        {TD \\ \scriptsize ($n=77$ across \\[-4pt] \scriptsize 12 storms)},
        {TS \\ \scriptsize ($n=605$ across \\[-4pt]  \scriptsize 29 storms)},
        {Category 1 \\ \scriptsize ($n=304$ across \\[-4pt]  \scriptsize 16 storms)},
        {Category 2 \\ \scriptsize ($n=153$ across \\[-4pt]  \scriptsize 7 storms)},
        {Category 3 \\ \scriptsize ($n=113$ across \\[-4pt]  \scriptsize 8 storms)},
        {Category 4 \\ \scriptsize ($n=117$ across \\[-4pt]  \scriptsize 6 storms)},
        {Category 5 \\ \scriptsize ($n=32$ across \\[-4pt]  \scriptsize 2 storms)},
    },
    x tick label style={align=center},
    xtick=data,
    ymajorgrids,
    yminorgrids,
    grid style={line width=.4pt, draw=gray!25},
    axis line style={line width=.5pt, draw=gray!60},
    tick style={draw=none},
    tick label style={font=\small},
    label style={font=\small},
    legend style={
    at={(0,1)}, anchor=north west,
    xshift=0.25cm, yshift=-0.25cm,
    legend columns=-1, draw=gray!60, fill=white, font=\small,
    /tikz/every even column/.append style={column sep=0.9em},
    },
    legend image code/.code={\draw[#1] (0cm,-0.1cm) rectangle (0.28cm,0.14cm);},
    error bars/y dir=both,
    error bars/y explicit,
    error bars/error bar style={line width=.5pt, black},
    error bars/error mark options={rotate=90, mark size=1.7pt, line width=.5pt, black},
]
\addplot[fill=white, draw=black] table[
    x=cat, y=aifsmean,
    y error plus  expr=\thisrow{aifsupper}-\thisrow{aifsmean},
    y error minus expr=\thisrow{aifsmean}-\thisrow{aifslower},
  ] {\results};
\addplot[fill=black!80, draw=black] table[
    x=cat, y=predmean,
    y error plus  expr=\thisrow{predupper}-\thisrow{predmean},
    y error minus expr=\thisrow{predmean}-\thisrow{predlower},
  ] {\results};
\addplot[fill=gray!50, draw=black] table[
    x=cat, y=fnv3mean,
    y error plus  expr=\thisrow{fnv3upper}-\thisrow{fnv3mean},
    y error minus expr=\thisrow{fnv3mean}-\thisrow{fnv3lower},
  ] {\results};
\addplot[pattern=north east lines, draw=black] table[
    x=cat, y=ofclmean,
    y error plus  expr=\thisrow{ofclupper}-\thisrow{ofclmean},
    y error minus expr=\thisrow{ofclmean}-\thisrow{ofcllower},
  ] {\results};
\legend{AIFS, AIFS-TC, FNV3, OFCL}
\end{axis}
\end{tikzpicture}
\caption{
    \textbf{Intensity forecast MAE by category.}
    Shows intensity MAE averaged over lead times 12--168 hours on the held-out year 2025 broken down by intensity category:
    tropical depression (TD; less than \SI{34}{kt}),
    tropical storm (TS; \qtyrange{34}{63}{kt}),
    Category 1 (\qtyrange{64}{82}{kt}),
    Category 2 (\qtyrange{83}{95}{kt}),
    Category 3 (\qtyrange{96}{112}{kt}),
    Category 4 (\qtyrange{113}{136}{kt}), and
    Category 5 (above \SI{136}{kt}).
    The numbers are computed from the subset of data points for storms in the North Atlantic and East Pacific for which forecasts are available for all models.
    Data points are categorised by the observed best-track intensity at forecast verification time, so a storm can contribute to multiple categories.
    Error bars indicate 95\% confidence intervals, which are computed by resampling according to a hierarchical procedure and taking the 2.5\% and 97.5\% quantiles:
    storms are resampled with replacement, and, for each storm, initialisation times are resampled with a circular block bootstrap with block length $\lceil \sqrt{n_{\text{init}}} \rceil$.
    The second level of sampling is included because data points in Category-5 belong to only two different storms, so resampling storms alone yields unrealistically narrow intervals for that category.
}
\label{fig:category}
\end{figure*}
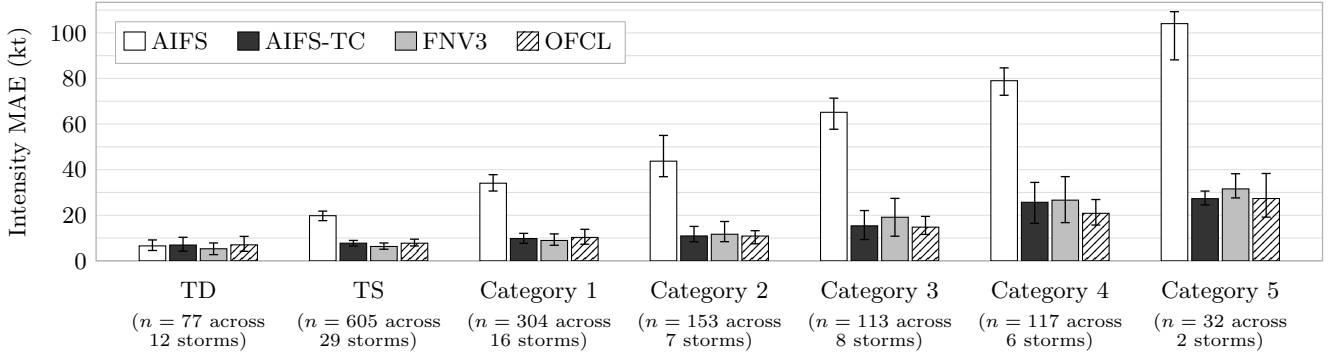

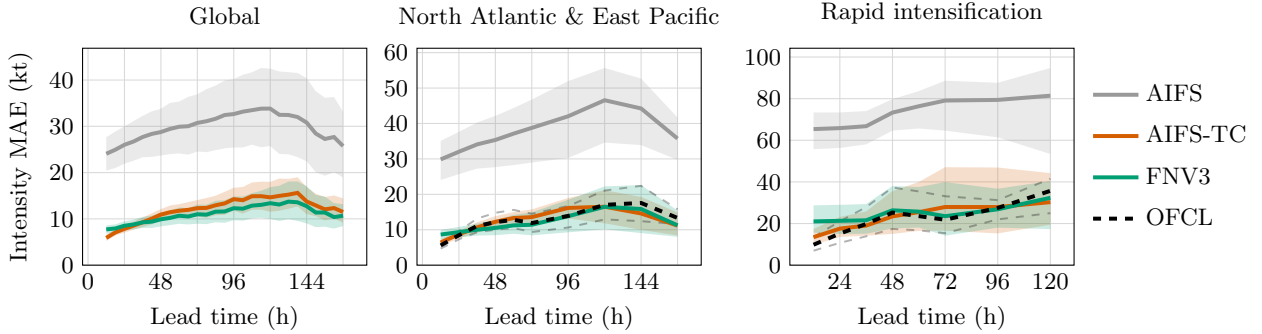
\begin{figure*}[t]
\small
\centering
\pgfplotstableread[col sep=comma]{
lead,aifsmean,aifslower,aifsupper,predmean,predlower,predupper,fnv3mean,fnv3lower,fnv3upper
12,24.07,20.42,27.64,5.89,5.23,6.57,7.73,6.82,8.57
18,24.95,21.00,28.87,7.07,6.39,7.72,7.93,7.02,8.81
24,26.01,21.94,30.01,7.85,7.03,8.60,8.52,7.45,9.48
30,26.72,22.56,31.13,8.54,7.65,9.41,8.80,7.75,9.84
36,27.69,23.32,32.27,9.26,8.21,10.29,9.25,8.07,10.44
42,28.36,23.40,33.23,10.05,8.81,11.21,9.40,8.04,10.69
48,28.77,23.69,33.80,10.93,9.58,12.12,9.91,8.38,11.36
54,29.44,23.98,35.06,11.44,9.95,12.88,10.21,8.66,11.79
60,29.91,24.15,35.35,11.76,10.27,13.19,10.68,8.94,12.45
66,30.02,23.87,36.36,11.95,10.22,13.63,10.51,8.80,12.34
72,30.74,24.55,36.93,12.32,10.54,14.06,11.00,8.95,13.08
78,31.10,24.41,38.03,12.37,10.23,14.52,10.91,8.86,13.06
84,31.63,24.86,38.79,12.94,10.78,15.01,11.56,9.27,13.90
90,32.44,25.23,39.89,13.34,10.86,15.85,11.60,9.16,14.02
96,32.65,25.43,40.22,14.26,11.80,16.73,12.31,9.71,14.86
102,33.17,25.42,41.13,14.03,11.13,16.89,12.20,9.61,14.84
108,33.49,26.03,41.68,14.89,11.96,17.87,12.86,10.10,15.58
114,33.82,25.47,42.61,14.91,11.99,18.05,13.03,10.24,16.08
120,33.83,25.74,42.46,14.65,11.55,18.00,13.39,9.96,16.86
126,32.48,24.02,41.86,14.97,11.75,18.72,13.10,9.54,17.16
132,32.45,23.95,41.70,15.24,12.19,18.65,13.73,9.93,18.08
138,32.01,23.35,41.53,15.62,12.68,19.00,13.58,9.56,18.10
144,30.81,22.44,40.40,13.76,11.03,16.88,12.75,9.02,17.07
150,28.44,20.09,37.59,12.91,10.58,15.44,11.37,8.21,15.17
156,27.27,19.32,36.17,12.06,9.67,14.75,11.38,8.31,14.82
162,27.69,20.08,36.33,12.34,10.08,15.07,10.39,7.90,13.41
168,25.73,18.97,33.14,11.49,9.16,14.44,10.68,8.45,13.18
}\dataleadtimeglobal
\pgfplotstableread[col sep=comma]{
lead,aifsmean,aifslower,aifsupper,predmean,predlower,predupper,fnv3mean,fnv3lower,fnv3upper,ofclmean,ofcllower,ofclupper
12,29.86,24.04,35.09,6.40,5.35,7.33,8.64,7.21,9.96,5.57,4.68,6.38
24,32.02,25.70,37.64,8.71,7.58,9.68,9.33,7.78,10.80,8.45,7.30,9.64
36,34.10,27.26,40.27,10.66,9.16,12.11,10.00,8.35,11.71,11.07,9.29,13.15
48,35.35,27.67,42.14,12.24,10.65,13.82,10.52,8.46,12.77,12.14,10.00,14.78
60,37.13,28.26,44.37,13.30,11.55,15.00,11.23,8.86,14.20,12.80,10.50,15.61
72,38.76,28.96,46.76,13.57,11.18,15.75,11.41,8.42,14.91,11.89,9.35,14.51
96,42.04,30.27,51.97,16.16,13.17,18.97,13.83,9.27,18.46,13.91,10.63,16.70
120,46.56,34.60,55.67,16.50,11.82,21.17,16.48,10.12,22.23,17.03,12.92,21.06
144,44.24,33.90,52.69,14.59,9.96,19.27,15.85,9.05,22.63,17.56,12.42,22.40
168,35.74,29.71,41.68,11.17,8.38,14.86,11.25,8.00,15.41,13.33,11.60,15.77
}\dataleadtimenaep
\pgfplotstableread[col sep=comma]{
lead,aifsmean,aifslower,aifsupper,predmean,predlower,predupper,fnv3mean,fnv3lower,fnv3upper,ofclmean,ofcllower,ofclupper
12,65.38,55.75,73.36,13.44,9.72,17.39,21.03,15.61,28.73,9.81,6.96,13.52
24,65.85,56.30,73.39,17.56,13.54,23.15,21.30,14.98,29.00,15.19,10.71,21.04
36,66.77,58.04,74.04,19.02,14.16,25.85,21.56,15.08,29.37,19.81,13.87,27.89
48,73.30,64.65,79.77,23.44,15.02,35.57,26.36,17.62,38.03,25.25,17.41,37.33
60,76.36,65.67,83.64,25.48,16.46,39.98,25.72,18.09,37.56,23.57,16.75,35.42
72,79.09,64.63,88.63,27.94,16.53,47.08,23.53,14.16,39.91,21.82,15.36,33.12
96,79.40,61.40,87.63,27.92,15.26,46.92,26.84,18.02,36.74,27.50,22.00,31.25
120,81.43,53.40,94.75,30.29,19.41,44.23,32.37,17.32,40.46,35.71,25.00,41.50
}\dataleadtimeri

\newcommand{\figwidth}{.3\linewidth}
\newcommand{\figheight}{.25\linewidth}
\newcommand{\uncertaintyband}[4][0.2]{%
    \addplot [draw=none, name path=#4U, forget plot] table [x=lead, y=#4upper] {#3};
    \addplot [draw=none, name path=#4L, forget plot] table [x=lead, y=#4lower] {#3};
    \addplot [fill=#2, fill opacity=#1, draw=none, forget plot]
        fill between [of=#4U and #4L];
}
\newcommand{\uncertaintybandnofill}[3]{%
    \addplot [#1, line width=0.75pt, opacity=0.3, forget plot, dashed] table [x=lead, y=#3upper] {#2};
    \addplot [#1, line width=0.75pt, opacity=0.3, forget plot, dashed] table [x=lead, y=#3lower] {#2};
}

\begin{tikzpicture}
\begin{axis}[
    width=\figwidth, height=\figheight,
    title={Global},
    title style={align=center},
    xlabel={Lead time (h)},
    ylabel={Intensity MAE (kt)},
    grid=major,
    tick style={draw=none},
    grid style={gray!30},
    legend cell align=left,
    legend style={at={(1.03,0.5)}, anchor=west, draw=none, row sep=4pt},
    every axis plot/.append style={line width=1.5pt},
    ytick distance=10,
    xtick distance=48,
    minor x tick num=1,
    xmajorgrids, xminorgrids,
    ymin=0,
]
    \uncertaintyband{cbGray}{\dataleadtimeglobal}{aifs}
    \uncertaintyband{cbVermillion}{\dataleadtimeglobal}{pred}
    \uncertaintyband{cbTeal}{\dataleadtimeglobal}{fnv3}
    \addplot [cbGray] table [x=lead, y=aifsmean] {\dataleadtimeglobal};
    \addplot [cbVermillion] table [x=lead, y=predmean] {\dataleadtimeglobal};
    \addplot [cbTeal] table [x=lead, y=fnv3mean] {\dataleadtimeglobal};
\end{axis}
\end{tikzpicture}
\begin{tikzpicture}
\begin{axis}[
    width=\figwidth, height=\figheight,
    title={North Atlantic \& East Pacific},
    title style={align=center},
    xlabel={Lead time (h)},
    grid=major,
    tick style={draw=none},
    grid style={gray!30},
    legend cell align=left,
    legend style={at={(1.03,0.5)}, anchor=west, draw=none, row sep=4pt},
    every axis plot/.append style={line width=1.5pt},
    ytick distance=10,
    xtick distance=48,
    minor x tick num=1,
    xmajorgrids, xminorgrids,
    ymin=0,
]
    \uncertaintyband{cbGray}{\dataleadtimenaep}{aifs}
    \uncertaintyband{cbVermillion}{\dataleadtimenaep}{pred}
    \uncertaintyband{cbTeal}{\dataleadtimenaep}{fnv3}
    \uncertaintybandnofill{black}{\dataleadtimenaep}{ofcl}
    \addplot [cbGray] table [x=lead, y=aifsmean] {\dataleadtimenaep};
    \addplot [cbVermillion] table [x=lead, y=predmean] {\dataleadtimenaep};
    \addplot [cbTeal] table [x=lead, y=fnv3mean] {\dataleadtimenaep};
    \addplot [black, dashed] table [x=lead, y=ofclmean]   {\dataleadtimenaep};
\end{axis}
\end{tikzpicture}
\begin{tikzpicture}
\begin{axis}[
    width=\figwidth, height=\figheight,
    title={Rapid intensification},
    title style={align=center},
    xlabel={Lead time (h)},
    grid=major,
    tick style={draw=none},
    grid style={gray!30},
    legend cell align=left,
    legend style={at={(1.03,0.5)}, anchor=west, draw=none, row sep=4pt},
    every axis plot/.append style={line width=1.5pt},
    ytick distance=20,
    xtick distance=24,
    minor x tick num=0,
    xmajorgrids, xminorgrids,
    ymin=0,
]
    \uncertaintyband{cbGray}{\dataleadtimeri}{aifs}
    \uncertaintyband{cbVermillion}{\dataleadtimeri}{pred}
    \uncertaintyband{cbTeal}{\dataleadtimeri}{fnv3}
    \uncertaintybandnofill{black}{\dataleadtimeri}{ofcl}
    \addplot [cbGray] table [x=lead, y=aifsmean] {\dataleadtimeri};
    \addplot [cbVermillion] table [x=lead, y=predmean] {\dataleadtimeri};
    \addplot [cbTeal] table [x=lead, y=fnv3mean] {\dataleadtimeri};
    \addplot [black, dashed] table [x=lead, y=ofclmean]  {\dataleadtimeri};
    \legend{AIFS, AIFS-TC, FNV3, OFCL}
\end{axis}
\end{tikzpicture}

\caption{
    \textbf{Intensity forecast MAE by lead time.}
    For each case and lead time, numbers are computed from the subset of data points for which forecasts are available for all models.
    In particular, the numbers for rapid intensification are restricted to the North Atlantic and East Pacific to enable a comparison with OFCL.
    Shaded regions (for AIFS, AIFS-TC, and FNV3) and transparent lines (for OFCL) indicate 95\% confidence intervals, which are computed by resampling entire storms with replacement and taking the 2.5\% and 97.5\% quantiles.
}
\label{fig:performance_by_leadtime}
\end{figure*}

\begin{table}[t]
    \caption{
        \textbf{Performance is maintained with operational initial conditions.}
        Shows intensity MAE averaged over lead times 12--168 hours on the held-out year 2025 for AIFS-TC with non-operational IBTrACS initial conditions and operational ATCF a-deck CARQ initial conditions.
        The MAE is very slightly higher for the operational initial conditions, but the increase is well within the error of the MAE.
        To line up with \cref{tab:main}, the MAE is computed over all storms in the North Atlantic and East Pacific for which OFCL forecasts are available. 
        95\% confidence intervals (CIs) are computed by resampling entire storms with replacement and taking the 2.5\% and 97.5\% quantiles.
    }
    \label{tab:operational}
    \small
    \centering
    \begin{tabular}{lcc}
         \toprule
          & MAE (kt) & 95\% CI \\ \midrule
         \multicolumn{3}{l}{\textsc{N.\ Atl. \& E.\ Pac.}} \\
         \quad IBTrACS & 11.0 & [9.3, 12.4] \\
         \quad CARQ & 11.1 & [9.2, 12.6] \\
        \addlinespace[5pt]
         \multicolumn{3}{l}{\textsc{Rapid intensification}} \\
         \quad IBTrACS & 20.9 & [14.8, 29.3] \\
         \quad CARQ & 21.2 & [14.3, 31.2] \\
         \bottomrule
    \end{tabular}
\end{table}

\begin{figure*}[t]
\small
\centering
\pgfplotstableread[col sep=comma]{
lead,aifs,pred,fnv3,best
0,27.00,26.25,29.00,29.00
6,26.00,26.68,26.90,29.00
12,27.00,28.50,27.60,29.00
18,28.00,31.31,31.10,29.00
24,29.00,34.20,36.30,39.00
30,30.00,37.68,39.30,39.00
36,33.00,44.11,47.20,39.00
42,34.00,48.62,55.90,45.00
48,38.00,54.74,66.60,54.00
54,39.00,64.45,79.00,60.00
60,42.00,72.33,106.20,64.00
66,44.00,81.44,122.90,70.00
72,46.00,87.38,136.50,80.00
78,47.00,94.94,133.90,89.00
84,50.00,99.78,139.50,93.00
90,52.00,114.53,142.10,109.00
96,56.00,118.15,137.00,109.00
102,57.00,122.60,128.10,115.00
108,57.00,116.01,116.70,99.00
114,56.00,116.56,107.80,93.00
120,46.00,85.28,93.90,84.00
126,45.00,74.06,85.70,70.00
132,45.00,69.19,75.70,70.00
138,44.00,71.43,69.60,64.00
144,43.00,70.87,68.20,64.00
150,42.00,69.39,65.20,54.00
156,41.00,66.88,60.80,51.00
162,41.00,62.71,56.60,51.00
168,37.00,57.53,51.70,45.00
174,35.00,45.85,46.90,45.00
186,34.00,43.07,39.30,35.00
192,35.00,45.13,35.50,35.00
198,33.00,44.21,36.40,35.00
}\datafirst
\pgfplotstableread[col sep=comma]{
lead,aifs,pred,fnv3,best
0,86.00,141.61,144.00,144.00
6,63.00,135.64,120.30,138.00
12,59.00,121.77,112.70,128.00
18,57.00,117.72,103.60,124.00
24,57.00,114.59,105.50,124.00
30,55.00,107.81,105.90,119.00
36,56.00,103.68,104.50,115.00
42,53.00,102.44,105.60,115.00
48,48.00,96.07,98.30,109.00
54,37.00,80.99,84.70,105.00
60,27.00,53.86,68.10,80.00
}\datasecond
\pgfplotstableread[col sep=comma]{
lead,aifs,pred,fnv3,best
0,49.00,120.04,124.00,125.00
6,43.00,126.08,132.50,135.00
12,47.00,127.74,127.20,145.00
18,43.00,127.49,128.80,155.00
24,43.00,128.27,131.40,155.00
30,47.00,129.93,128.40,155.00
36,46.00,130.18,129.40,165.00
42,44.00,119.87,118.60,145.00
48,44.00,114.97,109.40,105.00
54,49.00,118.25,108.70,105.00
60,46.00,107.54,94.90,85.00
66,37.00,98.42,94.90,80.00
72,42.00,87.68,94.30,80.00
78,43.00,86.83,101.50,85.00
84,46.00,88.97,104.60,90.00
90,48.00,88.72,102.30,85.00
96,50.00,85.27,94.40,80.00
102,52.00,85.25,84.10,75.00
108,52.00,82.16,74.80,75.00
114,50.00,73.47,68.20,70.00
}\datathird

\newcommand{\figwidth}{.3\linewidth}
\newcommand{\figheight}{.25\linewidth}

\begin{tikzpicture}
\begin{axis}[
    width=\figwidth, height=\figheight,
    title={Fung-Wong (WP)\\ Init.\ 2025-11-05 00Z},
    title style={align=center},
    xlabel={Lead time (h)},
    ylabel={Intensity (kt)},
    grid=major,
    tick style={draw=none},
    grid style={gray!30},
    legend cell align=left,
    legend style={at={(1.03,0.5)}, anchor=west, draw=none, row sep=4pt},
    every axis plot/.append style={line width=1.5pt},
    ytick distance=25,
    xtick distance=48,
]
    \addplot [cbGray] table [x=lead, y=aifs] {\datafirst};
    \addplot [cbVermillion] table [x=lead, y=pred] {\datafirst};
    \addplot [cbTeal] table [x=lead, y=fnv3] {\datafirst};
    \addplot [black, dashed] table [x=lead, y=best]   {\datafirst};
\end{axis}
\end{tikzpicture}
\begin{tikzpicture}
\begin{axis}[
    width=\figwidth, height=\figheight,
    title={Ragasa (WP)\\ Init.\ 2025-09-22 00Z},
    title style={align=center},
    xlabel={Lead time (h)},
    grid=major,
    tick style={draw=none},
    grid style={gray!30},
    legend cell align=left,
    legend style={at={(1.03,0.5)}, anchor=west, draw=none, row sep=4pt},
    every axis plot/.append style={line width=1.5pt},
    ytick distance=25,
    xtick distance=12,
]
    \addplot [cbGray] table [x=lead, y=aifs] {\datasecond};
    \addplot [cbVermillion] table [x=lead, y=pred] {\datasecond};
    \addplot [cbTeal] table [x=lead, y=fnv3] {\datasecond};
    \addplot [black, dashed] table [x=lead, y=best]   {\datasecond};
\end{axis}
\end{tikzpicture}
\begin{tikzpicture}
\begin{axis}[
    width=\figwidth, height=\figheight,
    title={Melissa (NA)\\ Init.\ 2025-10-27 00Z},
    title style={align=center},
    xlabel={Lead time (h)},
    grid=major,
    tick style={draw=none},
    grid style={gray!30},
    legend cell align=left,
    legend style={at={(1.03,0.5)}, anchor=west, draw=none, row sep=4pt},
    every axis plot/.append style={line width=1.5pt},
    ytick distance=25,
    xtick distance=24,
]
    \addplot [cbGray] table [x=lead, y=aifs] {\datathird};
    \addplot [cbVermillion] table [x=lead, y=pred] {\datathird};
    \addplot [cbTeal] table [x=lead, y=fnv3] {\datathird};
    \addplot [black, dashed] table [x=lead, y=best]   {\datathird};
    \legend{AIFS, AIFS-TC, FNV3, IBTrACS}
\end{axis}
\end{tikzpicture}

\caption{
    \textbf{Selected intensity forecasts.}
    Compares intensity forecasts from AIFS, AIFS-TC, and FNV3 against the best track from IBTrACS for two selected storms from the West Pacific (WP) and one from the North Atlantic (NA).
    The titles show the storm names.
}
\label{fig:cases}
\end{figure*}
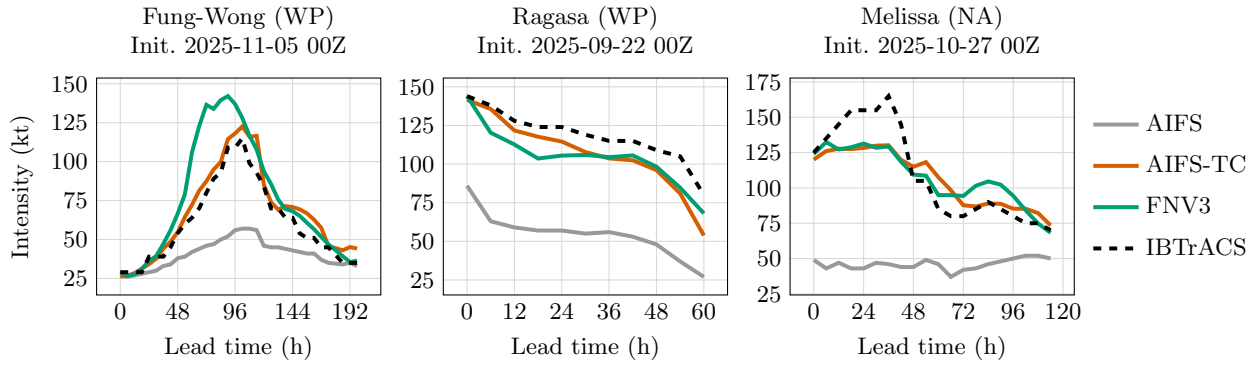

\section{Results: global and per-basin performance}
AIFS-TC reduces AIFS TC intensity forecast error by roughly a factor three and achieves performance competitive with the operational frontier (\cref{fig:headline,tab:main}).
Globally, across 70 storms in 2025, 
AIFS achieves an MAE of \SI{28.9}{kt}.
AIFS-TC is able to reduce this to \SI{10.8}{kt}, which is comparable to FNV3's \SI{10.2}{kt} (not statistically significantly different).
On the North Atlantic and East Pacific, where NHC
forecasts are also available, all three systems perform similarly: 11.0~kt for AIFS-TC, 10.7~kt for FNV3, and
10.7~kt for OFCL (pairwise not statistically significantly different).
For minimum central pressure, AIFS achieves an MAE of \SI{11.6}{mb} globally.
AIFS-TC reduces this to \SI{7.9}{mb}, which is comparable to FNV3's \SI{7.8}{mb} (not statistically significantly different).

The $\sim$$0.25^\circ$ grid of AIFS is too coarse to accurately represent the inner core of a TC.
Due to resolution and deterministic training methods \citep{philippe2026training}, raw AIFS forecasts blur the inner core and consequently severely underestimate intensity, leading to
a bias of \SI{-28.8}{kt} globally.
AIFS-TC reduces this bias to \SI{-1.9}{kt}, which is slightly smaller in magnitude than FNV3's \SI{-4.2}{kt} (statistically significant).
Similarly, for minimum central pressure, AIFS has a bias of \SI{10.2}{mb}.
AIFS-TC reduces this to \SI{1.5}{mb}, which is slightly smaller in magnitude than FNV3's \SI{3.4}{mb} (statistically significant).
If we stratify the forecast errors by ocean basin, the same trends hold:
the MAE and bias achieved by AIFS-TC are generally comparable to and not statistically different from those of FNV3.

For storms in the North Atlantic and East Pacific, \cref{fig:category} breaks down the intensity forecast MAE according to intensity category.
The MAE of AIFS-TC, FNV3, and OFCL is comparable for all categories;
differences are within the 95\% confidence intervals.
\Cref{fig:performance_by_leadtime} breaks the intensity forecast MAE by lead time.
AIFS-TC, FNV3, and OFCL (where available) again achieve similar MAE.
AIFS-TC attains a slightly lower MAE at short lead times and FNV3 at long lead times, but these differences again lie within the error bounds.

The results in \cref{tab:main} use IBTrACS initial conditions.
Although best-track data provide the most accurate estimates of track and intensity, they are determined retrospectively in post-storm analysis and are not available in real time.
\Cref{tab:operational} shows that replacing IBTrACS initial conditions with operationally available ones (ATCF a-deck CARQ rows) increases the MAE in the North Atlantic and East Pacific and in cases of rapid intensification by at most \SI{0.3}{kt}, well within the 95\% confidence intervals.

\section{Results: rapid intensification}
The forecasts that matter most, and that machine-learning models handle worst, are those of rapid
intensification (RI), conventionally defined as an increase of at least \SI{30}{kt} in maximum wind speed over \SI{24}{h}.
Raw AIFS forecasts completely fail to capture RI, resulting in an MAE of \SI{71.1}{kt} (\cref{fig:headline,tab:main}).
AIFS-TC is able to address this failure mode and brings the MAE down to \SI{20.9}{kt},
slightly lower than FNV3's \SI{23.8}{kt} (not statistically significant) and comparable to OFCL's \SI{20.1}{kt} (not statistically significantly different).
Similar results hold for pressure: \SI{18.8}{mb} for AIFS-TC versus \SI{20.8}{mb} for FNV3 (not statistically significantly different).

All systems, including human forecasters (OFCL), retain a substantial negative bias at RI, which reflects how hard the
timing of intensification is.

\section{Case studies}
\Cref{fig:cases} shows three selected intensity forecasts. In the first, Typhoon Fung-Wong peaked near \SI{115}{kt}.
FNV3 overestimates intensity by roughly \SI{25}{kt}, while AIFS-TC tracks the observed life cycle almost
exactly.
The second shows Category-5 Typhoon Ragasa.
The AIFS-TC forecast follows the slow post-peak decay that FNV3 loses.
The third is Hurricane Melissa near Jamaica, a violent Category-5 rapid intensifier that every system
under-forecasts at the peak.
Both AIFS-TC and FNV3 roughly capture the intensity evolution of Melissa through the mature phase.
In all three, raw AIFS (grey) never leaves the tropical-storm range, which is the failure mode that AIFS-TC is designed to repair.

\section{Relationship to prior work}
The previous approach closest to AIFS-TC is that of
\citet{gomez2025global}, who use a very similar setup.
Whereas they carefully investigate post-processing using Pangu-Weather and FourCastNet V2, we start from the stronger AIFS forecast system and learn a direct correction to its predicted intensity, rather than forecasting intensity changes relative to the observed initial state. Their framework deliberately avoids reliance on a tracker by using large regions centred on the initial storm position, while tracking is integrated directly into our pipeline, allowing us to use compact storm-centred patches and a richer set of physically derived predictors from a broader collection of AIFS fields.
More generally, the two studies also have different aims. Gomez et al.~compare linear, multilayer-perceptron and convolutional approaches in a controlled, ablation-style analysis designed to understand where predictive skill comes from. We instead combine complementary gradient-boosted and convolutional models to maximise forecast performance in a simple, inexpensive, operationally focussed system. Our study also uses a slightly larger and more recent training period. We therefore see their work as providing careful scientific evidence for the value of AI-weather-model post-processing, while ours focuses on translating that idea into a competitive operational forecasting system.

\section{Discussion and conclusion}
AIFS-TC is a simple correction to AIFS that is competitive with the operational frontier of TC intensity forecasting.
On the held-out year 2025, the MAE and bias of AIFS-TC for maximum wind speed and minimum central pressure are generally comparable to and not statistically significantly different from those of FNV3 and, where available, OFCL.
This holds even for cases of rapid intensification, which is the regime where AI weather models tend to fail most severely.
AIFS-TC therefore demonstrates that state-of-the-art TC intensity forecasting does not require a purpose-built system, but can currently be achieved by a simple machine-learning approach on top of an open AI weather model.

The approach can still be improved in various ways.
Training uses only the years 2016--2024, a single deterministic AIFS run, and no observational data.
Extending training to AIFS hindcasts further back;
incorporating the full AIFS ensemble to produce probabilistic intensity forecasts;
and augmenting the inputs to the GBM and CNN with observational data that has established intensity skill, such as satellite imagery and ocean heat content \citep{ships}, could all further improve performance.

A notable feature of this work is that AIFS-TC was produced by an LLM coding agent (Claude Fable 5) under the direction of a single domain scientist.
The scientist specified the goal and evaluation protocol and rigorously verified the outputs;
the agent implemented and trained the models and ran the evaluation.
For problems well matched to standard machine-learning techniques,
this shows that agentic coding can enable domain scientists to build and test systems competitive with the state of the art without specialist engineering or machine-learning support.

\paragraph{Data availability.}
TC best-track data is from IBTrACS~v04r01 (NOAA NCEI):
for maximum wind speed, we use \texttt{USA\_WIND} with \texttt{WMO\_WIND/0.88} as a fallback \citep{wmo2017guide}; and
for minimum central pressure, we use \texttt{USA\_PRES} with \texttt{WMO\_PRES} as a fallback.
FNV3 forecasts are from Google DeepMind WeatherLab.
Official
forecasts (OFCL) are from the NHC ATCF a-decks.

\paragraph{Code availability.}
The model and training pipeline are available at \url{https://github.com/cambridge-mlg/aifs-tc}.

\paragraph{Acknowledgements.}
Richard E. Turner is supported by the EPSRC Probabilistic AI Hub (EP/Y028783/1). Some of the computational work was funded by a Microsoft AI4Good Azure award.

\paragraph{Author contributions.} A.A.\ conceived and directed the study and provided scientific oversight;
C.F.5 (Claude~Fable~5) designed, implemented, trained and evaluated the system under A.A.'s natural-language direction;
M.M.-G.\ produced the AIFS historical cropped fields, AIFS TC track data, and contributed to evaluation;
H.C. assisted with AIFS forecasts and guidance on framing an operationally feasible system;
M.C.\ advised on AIFS and operational forecasting;
W.P.B.\ helped validate the results;
and W.P.B.\ and R.E.T.\ advised on methodology and framing. All authors reviewed the manuscript.

\bibliography{bibliography}

\end{document}